\documentclass[aps,chaos,floats,twocolumn,showpacs,superscriptaddress,reprint]{revtex4-1}

\usepackage{hyperref}
\usepackage{amsmath}
\usepackage{amssymb}
\usepackage{color}
\usepackage{graphics,graphicx}
\usepackage{pstricks,pst-node,pst-tree}

\newcommand{\Z}{\mathbb{Z}}

\newenvironment{weightedexample1}[6]
 {$
  \psmatrix[colsep=1.5cm, rowsep=0.25cm,mnode=circle]
      &  #1  &  #2    \\
  #3  &      &     &  #4  \\
      &  #5  &  #6
    \ncline{2,1}{1,2}
    \ncput*{1}
   \ncline{1,2}{1,3}
   \ncput*{1}
   \ncline{1,3}{2,4}
   \ncput*{1}
   \ncline{2,4}{3,3}
   \ncput*{1}
   \ncline{3,3}{3,2}
   \ncput*{1}
   \ncline{3,2}{2,1}
   \ncput*{1}
   \ncarc[arcangle=-60]{2,1}{2,4}
   \ncput*{1}
   \ncline{1,2}{3,2}
   \ncput*{3}
   \ncline{1,3}{3,3}
   \ncput*{1}
  \endpsmatrix
 $}

\newenvironment{weightedexample2}[6]
 {$
  \psmatrix[colsep=1.5cm, rowsep=0.25cm,mnode=circle]
      &  #1  &  #2    \\
  #3  &      &     &  #4  \\
      &  #5  &  #6
    \ncline{2,1}{1,2}
    \ncput*{2}
   \ncline{1,2}{1,3}
   \ncput*{1}
   \ncline{1,3}{2,4}
   \ncput*{2}
   \ncline{2,4}{3,3}
   \ncput*{2}
   \ncline{3,3}{3,2}
   \ncput*{2}
   \ncline{3,2}{2,1}
   \ncput*{2}
   \ncarc[arcangle=-60]{2,1}{2,4}
   \ncput*{2}
   \ncline{1,2}{3,2}
   \ncput*{1}
   \ncline{1,3}{3,3}
   \ncput*{2}
  \endpsmatrix
 $}

\newenvironment{graphexample}[6]
 {$
  \psmatrix[colsep=1.5cm, rowsep=0.25cm,mnode=circle]
      &  #1  &  #2    \\
  #3  &      &     &  #4  \\
      &  #5  &  #6
    \ncline{2,1}{1,2}
   \ncline{1,2}{1,3}
   \ncline{1,3}{2,4}
   \ncline{2,4}{3,3}
   \ncline{3,3}{3,2}
   \ncline{3,2}{2,1}
   \ncarc[arcangle=-60]{2,1}{2,4}
   \ncline{1,2}{3,2}
   \ncline{1,3}{3,3}
  \endpsmatrix
 $}

\begin{document}

\title{A new method for comparing rankings through complex networks: \\ Model and analysis of competitiveness of major European soccer leagues}

\author{Regino~Criado}
\affiliation{Department of Applied Mathematics, Rey Juan Carlos University, Madrid (Spain)}
\affiliation{Center for Biomedical Technology (CTB), Technical University of Madrid (Spain)}
\author{Esther~Garc\'{\i}a}
\affiliation{Department of Applied Mathematics, Rey Juan Carlos University, Madrid (Spain)}
\author{Francisco~Pedroche}
\affiliation{Institut de Matem\`{a}tica Multidisciplin\`{a}ria, Universitat Polit\`{e}cnica de Val\`{e}ncia, Valencia (Spain)}
\author{Miguel~Romance}
\affiliation{Department of Applied Mathematics, Rey Juan Carlos University, Madrid (Spain)}
\affiliation{Center for Biomedical Technology (CTB), Technical University of Madrid (Spain)}


\begin{abstract}
In this paper we show a new technique to analyze families of rankings. In particular we focus on sports rankings and, more precisely, on soccer leagues. We consider that two teams compete when they change their relative positions in consecutive rankings. This allows to define a graph by linking teams that compete. We show how to
use some structural properties of this {\em competitivity graph} to  measure to what extend the teams in a league compete. These structural properties are the mean degree, the mean strength and the clustering coefficient. We give a generalization of the Kendall's correlation coefficient to more than two rankings. We also show how to make a dynamic  analysis of a league and how to compare different leagues. We apply this technique to analyze the four major European soccer leagues: Bundesliga, Italian Lega, Spanish Liga, and Premier League. We compare our results with the classical analysis of sport ranking based on measures of
competitive balance.
\end{abstract}

\maketitle

{\bf An important feature of a sport competition is the uncertainty about the outcome.
Sport industry, governments and followers are interested in having  some degree of uncertainty about the competition. In the field of sport rankings the term ``competitive balance" attends to measure this degree of uncertainty about the result of a competition. A high competitiveness means that there is high uncertainty about the teams ranking. Classical measures of competitiveness are based on the ratio of wins of each team or other related measures. In this paper we show a new perspective by using techniques from complex networks. We show how to use an ad hoc graph, that we called ``competitivity graph", to give some measures of the competitiveness of a family of rankings. As an application we make a comparison of the four major European soccer leagues during
2011-12 season and 2012-13 season.}

\section{Introduction}\label{sec:intro}

When studying the ranking of a sport league (e.g., the final rank in a soccer league) one can use usual statistical measures such as the mean and the standard deviation. In fact,  this is usually done in sports ranking \cite{DoGo}. We are interested in the dynamic behaviour of a sport league and therefore we need to compare $r \geq 2$ rankings.

Studies on comparison of $r$ rankings can be traced back to the seminal paper of Kendall  \cite{KeBa39}, where the Kendall's concordance coefficient is defined. Previous works were focused on the correlation of only two rankings, see, e.g \cite{Kendall38} where the Kendall's correlation coefficient $\tau$ is defined.
We can distinguish three ways of comparing two rankings: a) to use a correlation coefficient (e.g, Spearman's $\rho$ or Kendall's $\tau$, \cite{DiGr}) b) to use a distance between the rankings (such as Spearman's footrule $D$, \cite{SeSaQu}, or other metrics  \cite{Fa06}) and c), to use Kendall's concordance coefficient (see \cite{Leg} for a review). In \cite{EmMa} the authors prove that a) and b) are equivalent. In this paper we use a new technique of comparing $r$ rankings by using some measures derived from graph theory. In more detail, given $r$ rankings we use a {\em competitivity graph} (CG). Some properties of this CG give us information about the {\em competitiveness} of the $r$ rankings. The theoretical definition of the CG and its relation to some well known objects in graph theory are studied in \cite{CrGaPeRo}. In this paper we focus on how to use the CG  to compare the four major European soccer leagues. In the field of analysis of sport leagues it is commonly used the term {\sl competitive balance} to denote the analysis of sport rankings, see \cite{DoGo} or \cite{Groo} for historical background. For some results concerning competitive balance in European soccer leagues, see  \cite{Paw}, \cite{Fed}, \cite{Goo}. In section \ref{sec:classic} we review in detail some classical measures of competitive balance applied to sport rankings.

Our interest in competitiveness comes from the concept of {\em Competitivity group}, defined in~\cite{Pe},
related with rankings of users in Social Networks based on the Personalized PageRank (PPR); see~\cite{Peijbc}, \cite{Pe12}  for details on how to use PPR in the field of Social Networks.  In~\cite{GaPeRo} we show a theoretical analysis of PPR, that gives insight about the concept of competitiveness, and we introduced the concept of {\em effective competitors}. This latter concept motivated the study of the competitors in the frame of complex networks theory.

The structure of the paper is the following. In section~\ref{sec:basic} we introduce the basic notation and the main structural properties of the {\em competitivity graph} that may be used to analyze the competitiveness of a family of graphs. For each measure introduced we indicate the meaning  of {\sl being more competitive}. In section~\ref{sec:soccer} we analyze the competitiveness of the four major European soccer leagues by using the
previous measures. We show how to make an evolutionary analysis of each league, and how to compare the competitiveness of the leagues. In section~\ref{sec:classic} we review some classical measures of competitiveness and we compute them for the four studied leagues.  In section~\ref{sec:conclusion} we give some conclusions about the use of the structural properties introduced and we highlight some interesting conclusions derived from the application of our methodology to the four soccer leagues analyzed.

\section{Comparing rankings and complex network analysis: A mathematical model} \label{sec:basic}

Given a set of elements ${\mathcal N}=\{1,\dots, n\}$ that we will call {\it nodes} we define a {\it ranking} $c$ of ${\mathcal N}$ as any bijection $c:{\mathcal N}\to {\mathcal N}$. We will write $\prec_{c} j$ when node $i$ appears first than node $j$ in the ranking $c$. Given a finite family ${\mathcal R}=\{c_1, c_2,\dots, c_r \}$ of rankings  we say that the pair of nodes $(i,j)\in {\mathcal N}$ {\it compete} if there exists $t\in\{1,2,\ldots, r-1\}$ such that $i$ and $j$ exchange their relative positions between consecutive rankings $c_{t}$ and $c_{t+1}$.  We define the {\it competitivity graph} of the family of rankings ${\mathcal R}$, denoted by $G_c({\mathcal R})=({\mathcal N}, E_{\mathcal R})$, where $E_{\mathcal R}$ denotes the set of edges, as the undirected graph with nodes ${\mathcal N}$ and edges  given by the rule: there is a link between $i$ and $j$ if $(i,j)$ compete.

For example, consider a set with 6 nodes, labeled from 1 to 6, and the following rankings:
\begin{align*}
c_1:\quad &(1,2,3,4,5,6),\\
c_2:\quad &(1,3,4,2,5,6),\\
c_3:\quad &(1,2,5,3,4,6),\\
c_4:\quad &(3,2,6,1,5,4),
\end{align*}
where the ranking position is given from left to right; for example, in ranking $c_4$ the node with highest position is node 3, the node with the second highest position is node 2, etc. The competitivity graph of this set of nodes and the family of rankings ${\mathcal R}=\{c_1,c_2,c_3,c_4\}$ is:

\bigskip

\centerline{\begin{graphexample}{2}{4}{1}{6}{3}{5}\end{graphexample}}

\bigskip\bigskip

\noindent There is a link between nodes 1 and 6 because they exchange their relative positions in rankings $c_3$ and $c_4$; similarly, there is a link between nodes 1 and 2 because node 1 appears first than node  2 in ranking $c_3$ but  node 2 appears first than node 1 when considering ranking $c_4$.

A deeper study of the structural properties of this type of graphs and the interplays with other classic families of graphs can be found in \cite{CrGaPeRo}.

We say that two nodes $i,j$ compete $k$-times if $k$ is the maximal number of rankings where $i$ and $j$ compete. The {\it evolutive competitivity graph} of ${\mathcal R}$, denoted by $G_c^e({\mathcal R})=({\mathcal N}, E_{\mathcal R}^e)$, will be the weighted undirected graph with nodes ${\mathcal N}$  and edges given by the rule: there is an edge between $i$ and $j$ labeled with weight $k$ if $(i,j)$ compete $k$ times. Note that the underlying (unweighed) network behind the (weighed) graph $G_c^e({\mathcal R})$ is $G_c({\mathcal R})$.

The evolutive competitivity graph $G^e_c({\mathcal R})$ of  the set of nodes $\{1,2,3,4,5,6\}$ and the family of rankings ${\mathcal R}=\{c_1,c_2,c_3,c_4\}$ is the weighted graph:

\bigskip
\centerline{\begin{weightedexample1}{2}{4}{1}{6}{3}{5}\end{weightedexample1}}
\bigskip
\bigskip\medskip

\noindent The link between nodes 2 and 3 has weight 3 because this pair of nodes compete 3 times: they exchange their relative position between rankings $c_1$ and $c_2$, between $c_2$ and $c_3$, and between $c_3$ and $c_4$. The rest of links are weighted with 1 because their extreme nodes only exchange their relative positions once.

The order of the rankings is fundamental in the calculation of the weights of the evolutive competitivity graph, although it does not have influence in the underlying (unweighted) competitivity graph. For example, if we now consider  the same set of nodes $\{1,2,3,4,5,6\}$ and the family of rankings $\hat{\mathcal R}=\{c_1,c_3,c_4,c_2\}$, we will obtain the following evolutive competitivity $G^e_c(\hat{\mathcal R})$ graph:

\bigskip
\centerline{\begin{weightedexample2}{2}{4}{1}{6}{3}{5}\end{weightedexample2}}
\bigskip
\bigskip\medskip

\noindent Notice that the list of node strengths in $G^e_c({\mathcal R})$ is $(3,5,5,3,3,3)$ while the list of node strengths for  $G^e_c(\hat{\mathcal R})$ is $(6,4,5,5,6,6)$.


\medskip

There are several ways to define the {\sl competitiveness} in order to compare two (or more) families of rankings ${\mathcal R}=\{c_1, c_2,\dots, c_r \}$ and ${\mathcal S}=\{c_1, c_2,\dots, c_s \}$ possibly coming from different sets of nodes or competitors ${\mathcal N}=\{1,\dots, n\}$ and ${\mathcal N'=\{1,\dots, n'\}}$. In the sequel, let $G_c^e({\mathcal R})=({\mathcal N}, E_{\mathcal R}^e)$ and $G_c^e({\mathcal S})=({\mathcal N'}, E_{\mathcal S}^e)$\ be two different evolutive competitivity graphs. Sometimes we will also refer to the underlying unweighted competitivity graphs  $G_c({\mathcal R})=({\mathcal N}, E_{\mathcal R})$ and $G_c({\mathcal S})=({\mathcal N'}, E_{\mathcal S})$.  We will consider, as measures of competitiveness, some different parameters:

\medskip
\noindent{\bf Normalized mean degree}.  We define the {\it normalized mean degree} of a family of rankings ${\mathcal R}$ as the sum of all the node degrees in the competitivity graph $G_c({\mathcal R})$ divided by the sum over all nodes of their highest possible degrees
\begin{equation}
{\rm ND}({\mathcal R})=\frac{1}{n(n-1)}\sum_{i\in{\mathcal N}} {\rm deg}(i).
\end{equation}
This parameter gives us a global idea about the relative number of times any two possible competitors have exchanged their respective positions through the corresponding family of rankings. In the previous example, the normalized mean degree ${\rm ND}({\mathcal R})=\frac{18}{5}$.

We say that {\it  ${\mathcal R}$ is more competitive than ${\mathcal S}$ with respect to the normalized mean degree} if $\displaystyle {\rm ND}({\mathcal R})>{\rm ND}({\mathcal S})$.

\medskip

\noindent{\bf Normalized mean strength}. The strength of a node in a weighted graph is the sum of the weights of its incident edges. We define the {\it normalized mean strength} of a family of rankings ${\mathcal R}$ as the sum of all edge weights in the evolutive competitivity graph $G_c^e({\mathcal R})$ divided by sum over all possible edges  of their highest  possible weights:
\begin{equation}
{\rm NS}({\mathcal R})=\frac{w(E_{\mathcal R}^e)}{\binom n 2(r-1)},
\end{equation}
where $w(E_{\mathcal R}^e)$ denotes the sum of all weights of the edges of the evolutive competitivity graph.

We say that {\it ${\mathcal R}$ is more competitive than ${\mathcal S}$ with respect to the normalized mean strength} if ${\rm NS}({\mathcal R})>{\rm NS}({\mathcal S})$.

In the examples above, $\hat{\mathcal R}$ is more competitive than ${\mathcal R}$ with respect to the normalized mean strength because  ${\rm NS}(\hat{\mathcal R})=\frac{16}{45}$ while ${\rm NS}({\mathcal R})=\frac{11}{45}$.

\medskip

\noindent{\bf Clustering coefficient}. In graph theory, a clique is a set of nodes mutually connected between them. For example,  a triangle is a clique formed by three nodes. The clustering coefficient measures how many nodes in a graph tend to cluster together.  The clustering coefficient $C_i$ of a node $i$ (see, for example \cite{Boccaletti06}) is defined as
\begin{equation}
C_i = \frac {e_i}{\binom{k_i}{ 2}},
\end{equation}
where $k_i$ is the number of neighbors of node $i$, $e_i$ is the number of connected pairs between the neighbors of $i$, and $\binom{k_i}{ 2}$ represents all possible pairs between the neighbors of $i$. Given a family of  rankings ${\mathcal R}$, the {\it clustering coefficient} of ${\mathcal R}$ is the average of the clustering coefficients of the nodes of the competitivity graph $G_c(\mathcal R)$, i.e.,
\begin{equation}
C({\mathcal R})=\frac 1 n \sum_{i\in {\mathcal N}} C_i.
\end{equation}
For instance, the clustering coefficient of ${\mathcal R}$ in the example above is $\frac 1 3$ since the clustering coefficient of each node of $G_c(\mathcal R)$ is $\frac 1 3$.

We say that {\it ${\mathcal R}$ is more competitive than ${\mathcal S}$ with respect to the clustering coefficient ${\mathcal C}$} if
$
{\mathcal C} ({\mathcal R}) > {\mathcal C}({\mathcal S})
$.


Similarly, we can consider other graph parameters such as the normalized size of the maximal clique (i.e., the number of nodes of the maximal clique contained in the graph divided by the number of nodes of the graph) the normalized size of the largest connected component, etc. For each of these parameters, a family of  rankings is more competitive than another family if this parameter in the (evolutive) competitivity graph is bigger, see \cite{suplementary}. There are other graph parameters that work the other way round: the smaller they are, the more competitive a family of  rankings is. Examples of such parameters are the number of connected components and the Kendall's coefficient $\tau$ that we will treat in the sequel.

\noindent {\bf Generalized Kendall's $\tau$ correlation coefficient}. It is well known that if we take two rankings $c_1$ and $c_2$ of a set $\mathcal{N}$ of $n$ elements, then the {\it Kendall's $\tau$ correlation coefficient} \cite{Kendall38} is defined as
\begin{equation}\label{kendalltau}
 \tau(c_1,c_2)=\frac {\tilde K(c_1,c_2)- K(c_1,c_2)}{\binom n 2},
\end{equation}
where $\tilde K(c_1,c_2)$ denotes the number of pairs $(i,j)$ that do not compete with respect to ${\mathcal R}=\{c_1,c_2\}$, and $K(c_1,c_2)$ denotes the number of pairs $(i,j)$ that  compete. Notice that ${\binom n 2}$ is the number of all possible pairs of nodes $(i,j)$. If we consider the competitivity graph $G_c({\mathcal R})$ with respect to ${\mathcal R}=\{c_1,c_2\}$ then $K(c_1,c_2)=|E_{\mathcal R}|$, the number of edges of $G_c({\mathcal R})$, and $\tilde K(c_1,c_2)=\binom n 2 -|E_{\mathcal R}|$, so
\begin{equation}
 \tau(c_1,c_2)=1-\frac {2|E_{\mathcal R}|}{\binom n 2}=1-\frac {4 |E_{\mathcal R}|}{n(n-1)}.
\end{equation}

The number of edges $E_{\mathcal R}$ of the competitivity graph $G_c({\mathcal R})$ of a family ${\mathcal R}$ of rankings is related with the Kendall's correlation coefficient of two rankings:   if we denote by   $E(c_1,c_2)$ the  edges of the competitivity graph of  the family $\{c_1,c_2\}$, we have that

\begin{align}
|E_{\mathcal R}| & \ge   \max_{c_1,c_2\in {\mathcal R}} |E(c_1,c_2)| \notag \\
                 & \ge \frac{n(n-1)}{4}\left(1-\min \tau(c_1,c_2)\right)
\end{align}
and this inequality is in fact an equality when $r=2$. Similarly, $E_{\mathcal R}=\bigcup_{c_1,c_2\in{\mathcal R}} E(c_1,c_2)$ so
\begin{align}
|E_{\mathcal R}|&\le \sum_{c_1,c_2\in {\mathcal R}}|E(c_1,c_2)|\notag \\
 &=  \frac{n(n-1)}{4} \left(\binom{r}{2}-\sum_{c_1,c_2\in {\mathcal R}} \tau(c_1,c_2) \right).
\end{align}
This inequality turns to be an equality when $r=2$.

We can define a {\it generalized  Kendall's  correlation coefficient} $\tau({\mathcal R})$ of a family ${\mathcal R}$ of $r\ge 2$ rankings: following the original definition (number of pairs that do not compete $\tilde K({\mathcal R})$ minus number of pairs that compete $K({\mathcal R})$, divided by the number of all possible pairs $\binom n 2$), we set
\begin{align}
 \tau({\mathcal R})&=\frac {\tilde K({\mathcal R})- K({\mathcal R})}{\binom n 2}=1-\frac {2|E_{\mathcal R}|}{\binom n 2}  \notag \\
 &=1-\frac {4 |E_{\mathcal R}|}{n(n-1)}.
\end{align}

We can also construct an {\it evolutive Kendall's correlation coefficient} $\tau({\mathcal R})_e$ if we take into account the number of times each pair of nodes compete. In this sense, we define
\begin{equation}
\tau({\mathcal R})_e=1-\frac {2\ w(E_{\mathcal R}^e)}{\binom n 2 (r-1)},
\end{equation}
where $w(E_{\mathcal R}^e)$ denotes the sum of all weights of the edges of the evolutive competitivity graph. The denominator $ \binom n 2 (r-1)$ represents the sum over all possible edges  of their highest possible weights.

The evolutive Kendall's coefficient of a family of rankings ${\mathcal R}$ is directly related to the normalized strength of the family
\begin{equation}
\tau_e({\mathcal R})=1-2\, {\rm NS}({\mathcal R}).
\end{equation}



We say that {\it ${\mathcal R}$ is more competitive than ${\mathcal S}$ with respect to the Kendall's coefficient} if
$ \tau_e({\mathcal R}) <  \tau_e({\mathcal S})$.
Notice that the smaller the Kendall's coefficient $\tau_e({\mathcal R})$ is, the more competitive ${\mathcal R}$ is.

\medskip
By using these parameters we can compare the competitiveness of the major European soccer leagues in 2011-12 and 2012-13. This is the main goal of next section.

\section{An application to the competitiveness of the Major European Soccer Leagues in 2011-12}\label{sec:soccer}

In this section we will illustrate the method introduced in section~\ref{sec:basic} by analyzing the competitiveness of four major European soccer leagues during 2011-12 season  and 2012-13 season.  From a competitiveness point of view, a season of a (soccer) league of $n$ teams is a family of $2n-2$ rankings ${\mathcal R}=\{c_1,\cdots, c_{2n-2}\}$ such that each $c_k$ corresponds to the classification of the teams at matchday $k$. Note that the ranking are temporarily ordered and they aggregate all the information about all the matches and their results from the beginning of the season until the corresponding matchday.

We have chosen the classifications of the following major European soccer leagues along 2011-12 and 2012-13:
\begin{itemize}
 \item The German {\bf Fu{\ss}ball-Bundesliga} \cite{web-bundes} (also called, in short, the {\sl Bundesliga}) which is the top league at the German soccer league system. During season 2011-12 and 2012-13 it is contested by 18 teams and operates on a system of promotion and relegation with the {\sl 2nd. Bundesliga}.
 \item The Italian {\bf Lega Serie A} \cite{web-calcio} (also called {\sl Serie A TIM} or the {\sl Calcio}, until 2010), which is at the top of the Italian soccer league system and it is contested by 20 Italian teams.
 \item The Spanish {\bf Liga BBVA} \cite{web-liga} (also called {\sl La Liga}), which is the First Division of the Spanish Professional Soccer National Association (LFP). It is contested by 20 teams, with the three lowest placed teams relegated to the {\sl Segunda Divisi\'on} and replaced by the top two teams in that division plus the winner of a play-off.
 \item The British {\bf Barclays Premier League} \cite{web-premier} (also simply called the {\sl Premier League})  which is the professional league at the top of the English men's association soccer clubs. It is contested by 20 clubs and it operates on a system of promotion and relegation with the {\sl Football League}.
\end{itemize}

\begin{figure*}
 \hfill
 \includegraphics[width=0.4\textwidth]{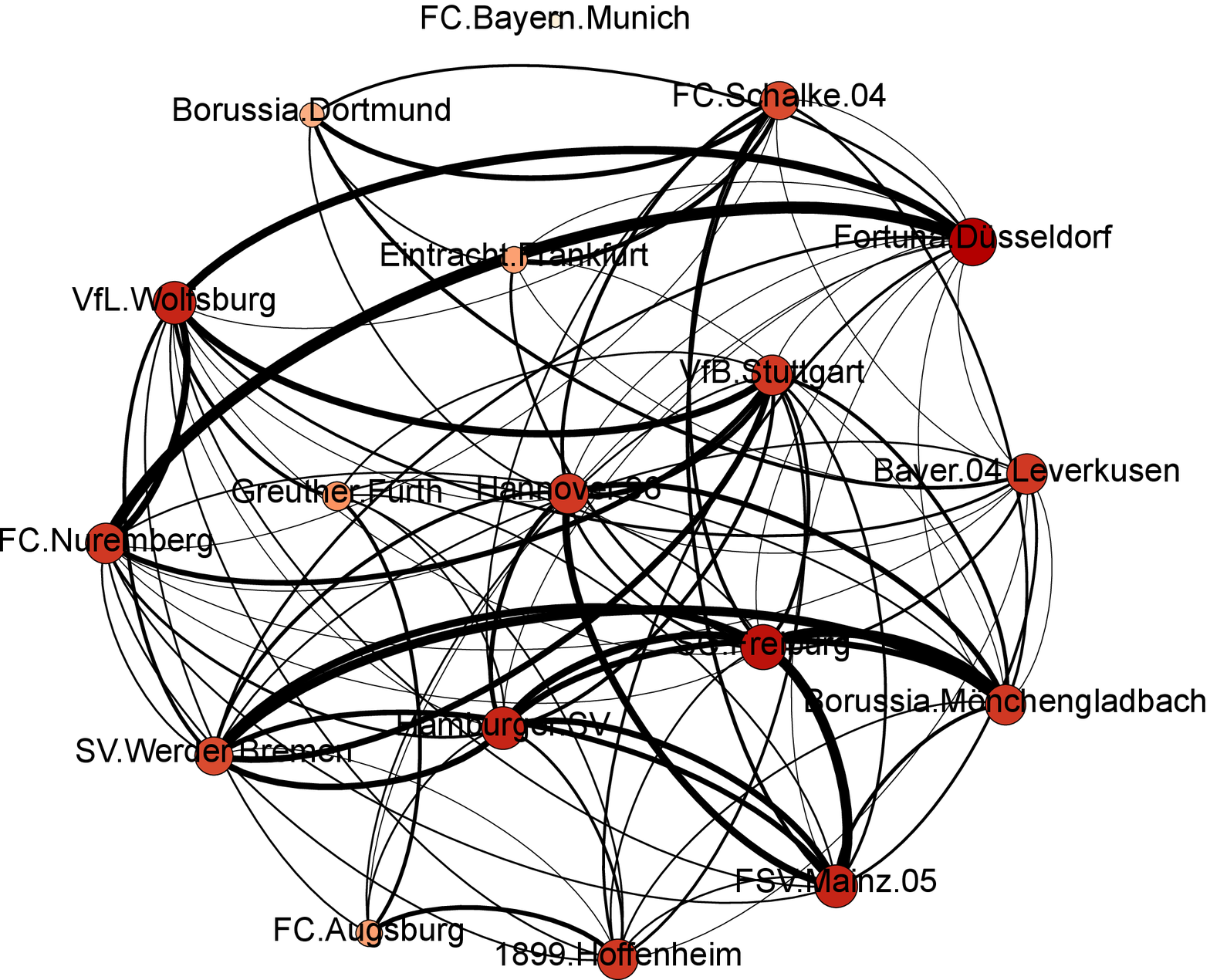}
 \hfill
 \includegraphics[width=0.4\textwidth]{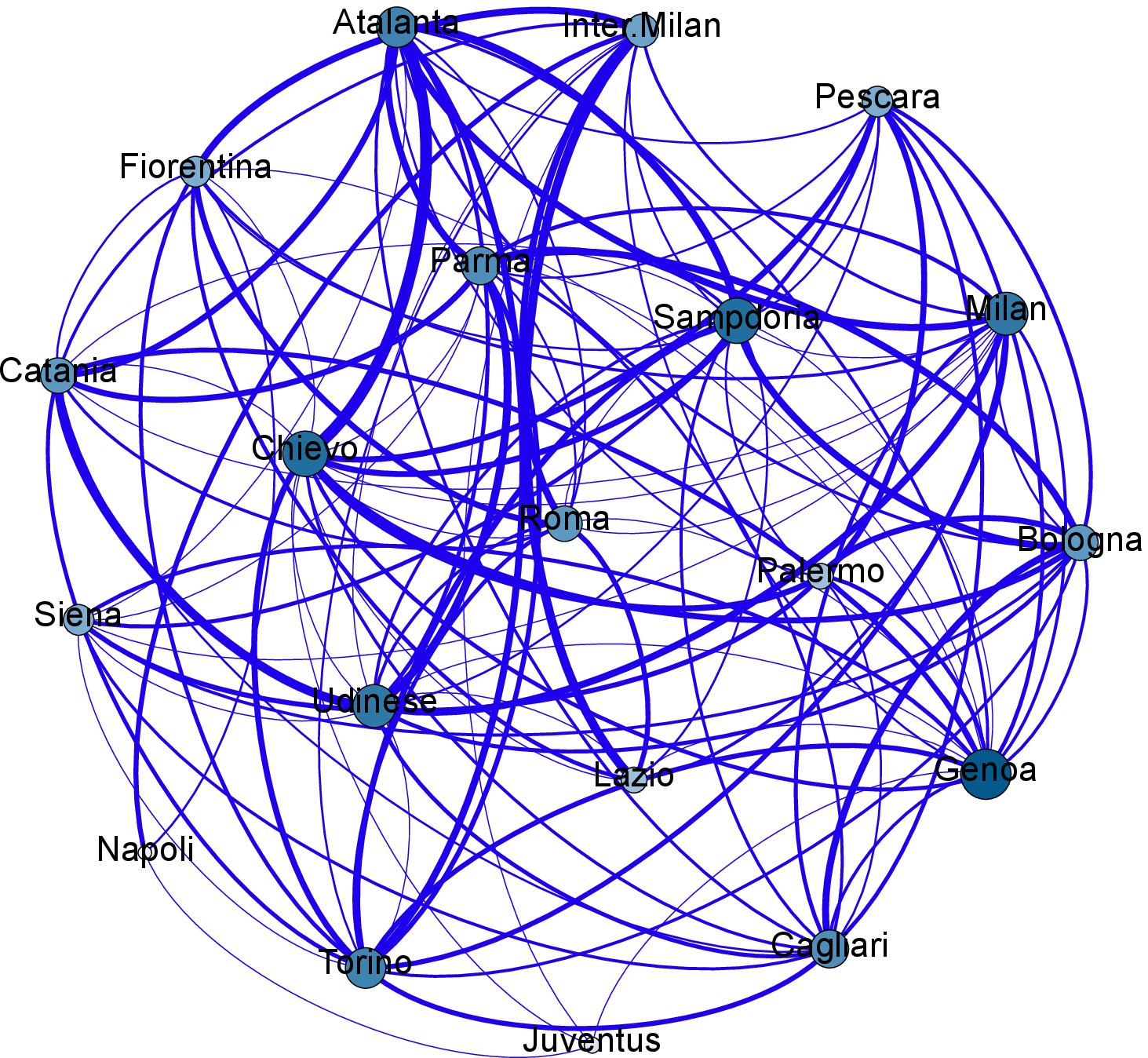}
 \hfill $\,$\hfill\\
 $\,$\\
 \hfill
 \includegraphics[width=0.4\textwidth]{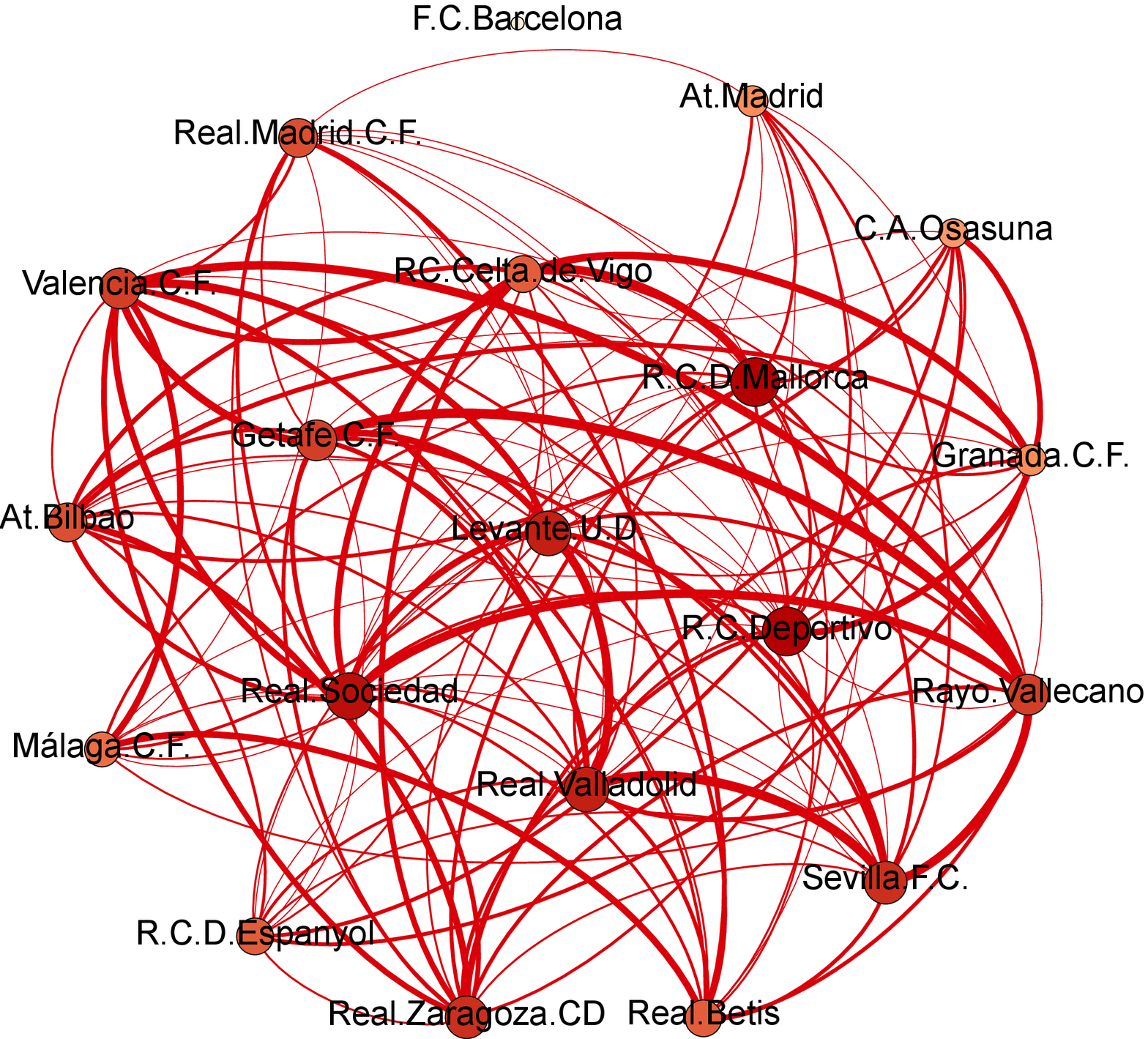}
 \hfill
 \includegraphics[width=0.4\textwidth]{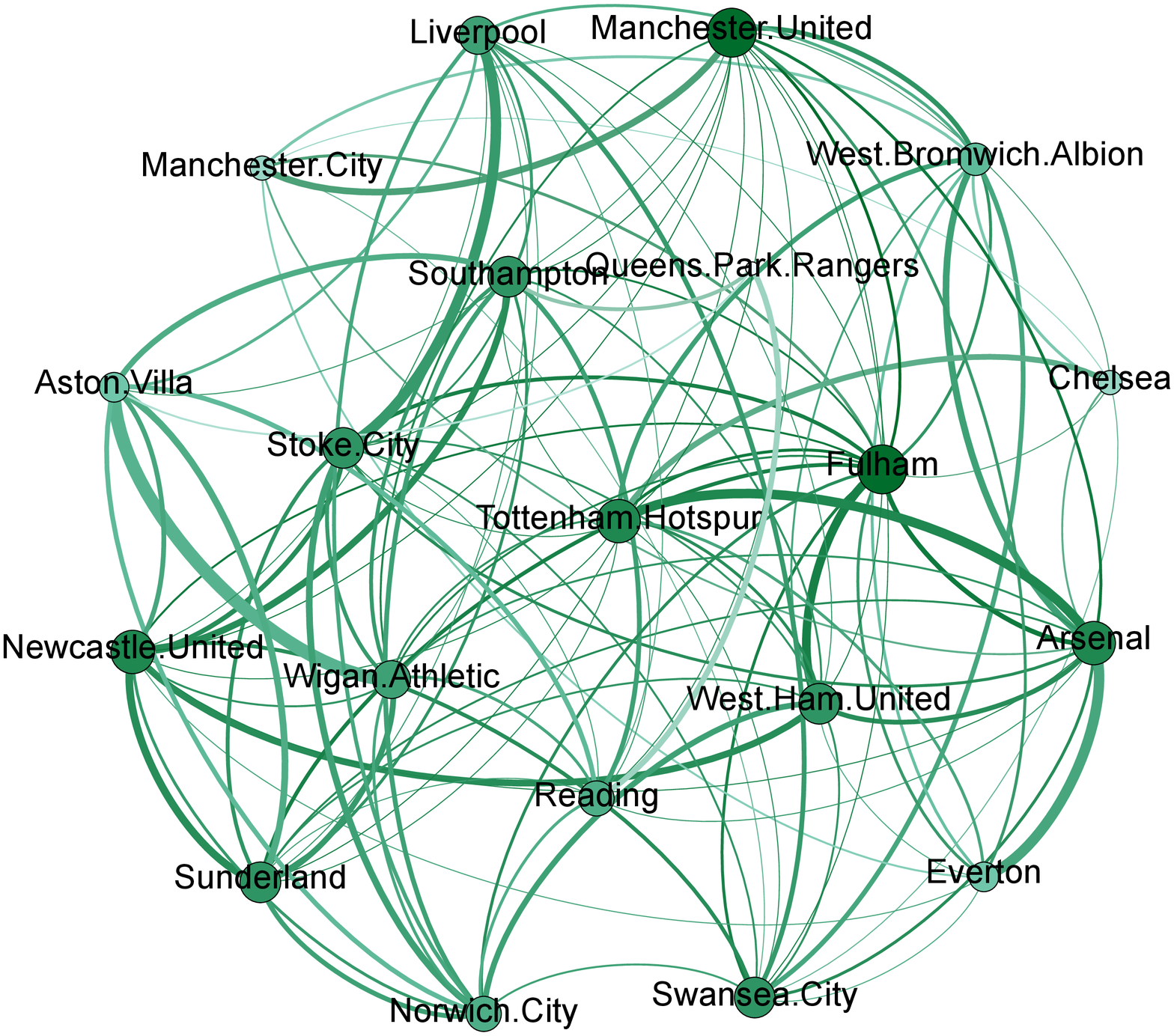}
 \hfill $\,$\hfill
\caption{\label{fig:networks}The evolutive competitivity graph for the German {\sl Bundesliga} (on black), the Italian {\sl Lega Serie A} (on blue), the Spanish {\sl Liga BBVA} (on red) and the British {\sl Premier} (on green) during season 2012-13.}
\end{figure*}

Since each National League of $n$ teams along a season can be understood as a (ordered) family ${\mathcal R}$ of $2n-2$ rankings, we can consider its (evolutive) competitivity graph $G^e_c({\mathcal R})$, that we will call the (evolutive) competitivity graph of the league in the corresponding season. Figure~\ref{fig:networks} shows the evolutive competitive graphs of the the four European Leagues considered during 2012-13 (see \cite{suplementary} for the competitivity graphs of the Leagues considered during 2011-12).
This Figure shows, e.g,  that the FC Bayern-M$\ddot{u}$nchen is not connected with any team: it occupied the first position
during all the season 2012-13. The same happens with FC Barcelona in the Spanish league.

Now we can obtain information about the competitiveness in each league from structural properties of the corresponding (evolutive) competitivity graph $G^e_c({\mathcal R})$, following the ideas introduced in section~\ref{sec:basic}. Table~\ref{table:data12-13} shows several structural parameters for the four Leagues during 2011-12 and 2012-13 season and we can see that there are some clear differences in the competitiveness among the leagues. Actually, some leagues are more competitive with respect to some parameters, while the ordering changes if we consider other structural measures. For example, the Bundesliga is more competitive than the Lega Serie A during 2011-12 season with respect to the Kendall's coefficient $\tau$, but  the Lega Serie A is more competitive than the Bundesliga in the same season with respect to the evolutive Kendall's coefficient $\tau_e$. In general, the results derived from the weight distributions of the (evolutive) competitivity graph (such as the normalized mean strength or the evolutive Kendall's coefficient $\tau_e$) are more accurate than those coming from the (unweighed) competitivity graph (such as the normalized mean degree). Following this idea, we get that during 2011-12 season the Italian Lega Serie A was the most competitive one, followed by the Spanish Liga BBVA, the German Bundesliga and finally the British Premier League. Similarly, during 2012-13 season, the most competitive league was the German Bundesliga, followed by the Spanish Liga BBVA, the Italian Lega Serie A and finally the British Premier League, despite the fact that either in the German and the Spanish Leagues the champion of each league (The Bayern M\"unchen and F.C.Barcelona) occupied the first position along the whole season.

\begin{table*}
\caption{\label{table:data12-13} Computation of the normalized mean degree (ND$({\mathcal R})$), the normalized mean strength (NS$({\mathcal R})$), the clustering coefficient (C$({\mathcal R})$), the generalized Kendall's $\tau$ coefficient ($\tau({\mathcal R})$), the evolutive Kendall's coefficient ($\tau_e({\mathcal R})$), the number of connected components (NC$({\mathcal R})$), the normalized size of the maximal connected component (SC$({\mathcal R})$) and the normalized size of the maximal clique (SQ$({\mathcal R})$) of the evolutive competitivity graph $G^e_v({\mathcal R})$ of the considered leagues in 2011-12 and 2012-13.}
$\,$

\begin{tabular}{|c|c|c|c|c|c|c|c|c|}
\hline
League and Season & ND$({\mathcal R})$ & NS$({\mathcal R})$ & C$({\mathcal R})$ & $\tau({\mathcal R})$ & $\tau_e({\mathcal R})$& NC$({\mathcal R})$ & SC$({\mathcal R})$ & SQ$({\mathcal R})$ \\ \hline
Bundesliga 2011-12 & 0.7255 & 0.0604 & 0.8126 & -0.4510 & 0.8792 & 1      & 1      & 0.5556 \\
Lega Serie A 2011-12     & 0.6895 & 0.0690 & 0.8300 & -0.3789 & 0.8620 & 1      & 1      & 0.5500 \\
Liga BBVA 2011-12  & 0.6368 & 0.0656 & 0.8273 & -0.2737 & 0.8688 & 1      & 1      & 0.5500 \\
Premier League 2011-12   & 0.5684 & 0.0572 & 0.7887 & -0.1368 & 0.8856 & 2      & 0.9000 & 0.6000 \\ \hline
Bundesliga 2012-13 & 0.6078 & 0.0636 & 0.7721 & -0.2157 & 0.8728 & 2      & 0.9444 & 0.5000 \\
Lega Serie A 2012-13     & 0.6158 & 0.0576 & 0.7801 & -0.2316 & 0.8848 & 1      & 1      & 0.5000 \\
Liga BBVA 2012-13  & 0.6737 & 0.0619 & 0.8143 & -0.3474 & 0.8762 & 2      & 0.9500 & 0.5500 \\
Premier League 2012-13   & 0.6053 & 0.0536 & 0.7710 & -0.2105 & 0.8927 & 1      & 1      & 0.4000 \\ \hline
\end{tabular}
\end{table*}

In addition to the {\sl static} analysis of certain structural parameters presented before, we can consider the evolution of these parameters along the season in order to measure the fluctuations of competitiveness. In fact we can associate a temporal (weighted) network $G^e_c({\mathcal R},t)$ to each league of $n$ teams as follows. For every $1\le t\le 2n-2$, $G^e_c({\mathcal R},t)=G^e_c({\mathcal R}_t)$, where ${\mathcal R}_t=\{c_1,\cdots, c_t\}$ (i.e. ${\mathcal R}_t$ is the set of rankings corresponding to the $t$ first matchdays). Hence, the structural evolution of some structural parameters of this time-varying topology network $G^e_c({\mathcal R},t)$ is useful for  analyzing the fluctuations of the competitiveness along the season. Furthermore, this evolution can help spotting the matchday that gave a boost in the competiteness of the league. In order to avoid spurious fluctuations coming from the first matchdays, we only analyze the evolution of the temporal network $G^e_c({\mathcal R},t)$ with $4\le t\le 2n-2$.

\begin{figure*}
 \includegraphics[width=\textwidth]{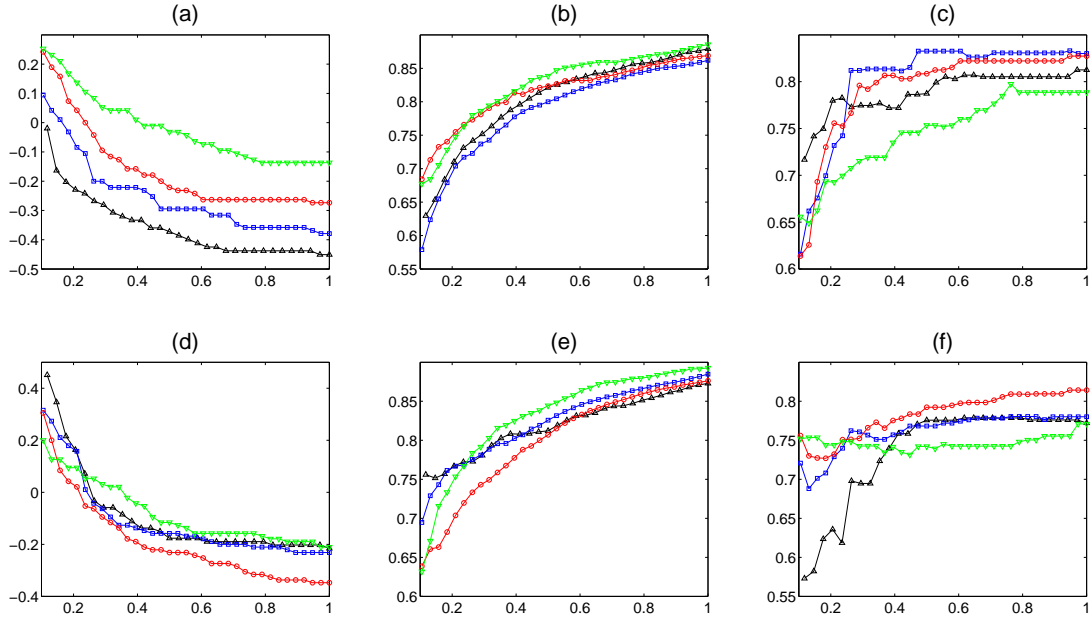}\\
 \caption{\label{fig:evolution}A comparison of the generalized Kendall's coefficient $\tau$ (panels (a) and (d)), the evolutive Kendall's coefficient $\tau_e$ (panels (b) and (e)) and the clustering coefficient (panels (c) and (f)) for the German Bundesliga (in black), the Italian Lega Serie A (in blue), the Spanish Liga BBVA (in red) and British Premier League (in green) during 2011-12 (first row, i.e. panels (a), (b) and (c)) and 2012-13 (second row, i.e. panels (d), (e) and (f)).}
\end{figure*}

Following this idea, Figure~\ref{fig:evolution} shows the evolution of the generalized Kendall's coefficient $\tau$ (see panels (a) and (d)), the evolutive Kendall's coefficient $\tau_e$ (panels (b) and (e)) and the clustering coefficient (panels (c) and (f)) for the four major European leagues considered during 2011-12 and 2012-13. This Figure shows that in both seasons the {\sl competitiveness rankings} among the four leagues are different if we consider generalized Kendall's coefficient $\tau$ or the evolutive Kendall's coefficient $\tau_e$. Furthermore, if we consider other structural parameter (the clustering coefficient), the corresponding ranking coincides with the ranking coming from the  evolutive Kendall's coefficient $\tau_e$ in 2011-12 season (where the Lega Serie A was the most competitive, followed by the Liga BBVA, the Bundesliga and finally the Premier League) and it coincides with the ranking coming from the  generalized Kendall's coefficient $\tau$ in 2012-13 season (where the most competitive league was the Spanish Liga BBVA, followed by Lega Serie A, the Bundesliga and the Premier League).

The analysis of the evolution of these structural parameters is also useful for studying the differences in the evolution of the competitiveness during 2011-12 and 2012-13. In 2011-12 season there were no significant fluctuations in the competitiveness rankings along the seasons (i.e. the competitiveness ranking among the four Leagues does not change too much along the season), while  in 2012-13 season there were some changes in the competitiveness ranking specially during the second half of the season. For example, panel (e) shows that, despite the fact that the competitiveness of the Bundesliga was quite low at the beginning of the season, it got better, by overcoming the BBVA Liga about matchday 20, which makes the Bundesliga more competitive since then. Similar analysis can be performed by considering other structural parameters (such as the normalized mean strength, the normalized size of the maximal clique and other), as it is illustrated in \cite{suplementary}.

\begin{figure*}
 \includegraphics[width=\textwidth]{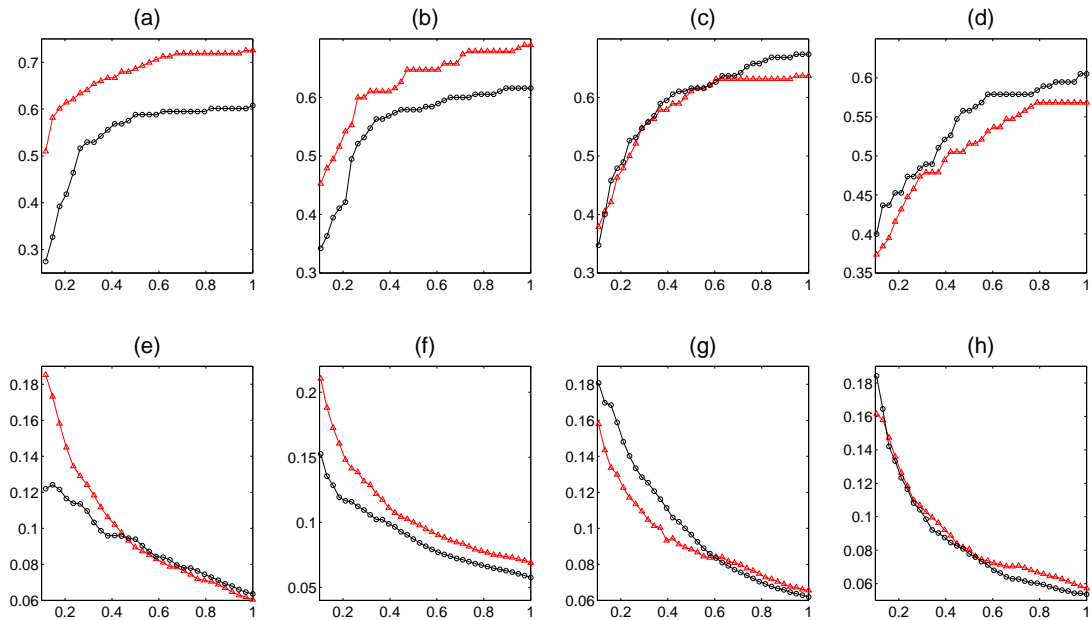}\\
 \caption{\label{fig:evolution02}A comparison of the normalized mean degree (panels from (a) to (d)) and the normalized mean strength (panels from (e) to (h)) for the German Bundesliga (panels (a) and (e)), the Italian Lega Serie A (panels (b) and (f)), the Spanish Liga BBVA (panels (c) and (g)) and British Premier League (panels (d) and (h)) during 2011-12 (lines in red in all the panels) and 2012-13 (lines in black in all the panels).}
\end{figure*}

Similarly, we can fix a league and compare the competitiveness of this league along different seasons  by analyzing the evolution of some structural parameters of the temporal network $G^e_c({\mathcal R},t)$ along different seasons. Figure~\ref{fig:evolution02} shows a comparison of the normalized mean degree (panels from (a) to (d)) and the normalized mean strength (panels from (e) to (h)) for the German Bundesliga (panels (a) and (e)), the Italian Lega Serie A (panels (b) and (f)), the Spanish Liga BBVA (panels (c) and (g)) and British Premier League (panels (d) and (h)) during 2011-12 season (lines in red in all the panels) and 2012-13 (lines in black in all the panels). This figure illustrates that the conclusions obtained strongly depend on the kind of structural parameter that we are considering. In general,  the parameters that take into account the weights of the evolutive competitivity graph are more sensible and therefore they give a better knowledge of the competitivenss of the soccer leagues.

In any case, the behavior of each of the four soccer leagues along 2011-12 and 2012-13 are quite different. For example, if we analyze the evolution of the normalized mean strength (panels from (e) to (h) in Figure~\ref{fig:evolution02}) we can see four different situations. The Bundesliga (panel (e)) was less competitive during the first half of 2011-12 than during the same period of 2012-13, but its competitiveness improved significantly in the second half of 2012-13 and it was more competitive during this period than in the previous season. The Lega Serie A (panel (b)) was definitively less competitive during 2012-13 than during 2011-12. The Liga BBVA exhibits a behavior that it is the opposite to the Bundesliga's one: it was much more competitive during the first half of 2012-13 season, but in the second half of the season the competitiveness was reduced and actually it was more competitive in the second half of 2011-12 than in 2012-13. Finally, the Premier League kept its competitiveness at the same level during 2011-12 and 2012-13. Similar analysis can be performed by considering other structural parameters (such as the normalized mean strength, the normalized size of the maximal clique and other), as it is illustrated in \cite{suplementary}.

We also consider another type of analysis of the competitiveness of the European soccer leagues that take into account the distribution of the weights of the evolutive competitivity graph. If we have two different leagues with the same strength at each node, the competitiveness should be different if these strengths are concentrated in some couples of nodes than if they are distributed in a big set of pairs of {\sl loosely competitor} nodes. In other words, we should measure how deeply the couples of teams compete in order to get a sharp idea of the competitiveness in each league. It should be interesting to take the evolutive competitivity graph of a league of a given season and only considering the competitiveness between nodes that flip their positions a significant number of times. In order to do this, we can fix a threshold $0\le \omega_o\in \Z$ and study the structural properties of the {\sl filtered} evolutive competitivity graph $G^e_c({\mathcal R})|_{\ge \omega_o}$, which is the graph whose (weighted) adjacency matrix $\tilde A=(\tilde a_{ij})$ is given by
\begin{equation}
\tilde a_{ij}=\left\{
         \begin{array}{ll}
           a_{ij}, & \hbox{if $a_{ij}\ge w_o$,} \\
           0, & \hbox{otherwise,}
         \end{array}
       \right.
\end{equation}
where $A=(a_{ij})$ is the (weighted) adjacency matrix of the evolutive competitivity graph $G^e_c({\mathcal R})$. Note that for every $0\le \omega_o\in\Z$, the {\sl filtered} evolutive competitivity graph $G^e_c({\mathcal R})|_{\ge \omega_o}$ only takes into account the competitiveness between the nodes such that compete at least $w_o$ times. By using this new tool, we can study how deeply the couples of teams compete in  a league by measuring several properties of $G^e_c({\mathcal R})|_{\ge \omega_o}$ and how they change when we consider different values of $0\le \omega_o\in\Z$.

\begin{figure*}
 \includegraphics[width=\textwidth]{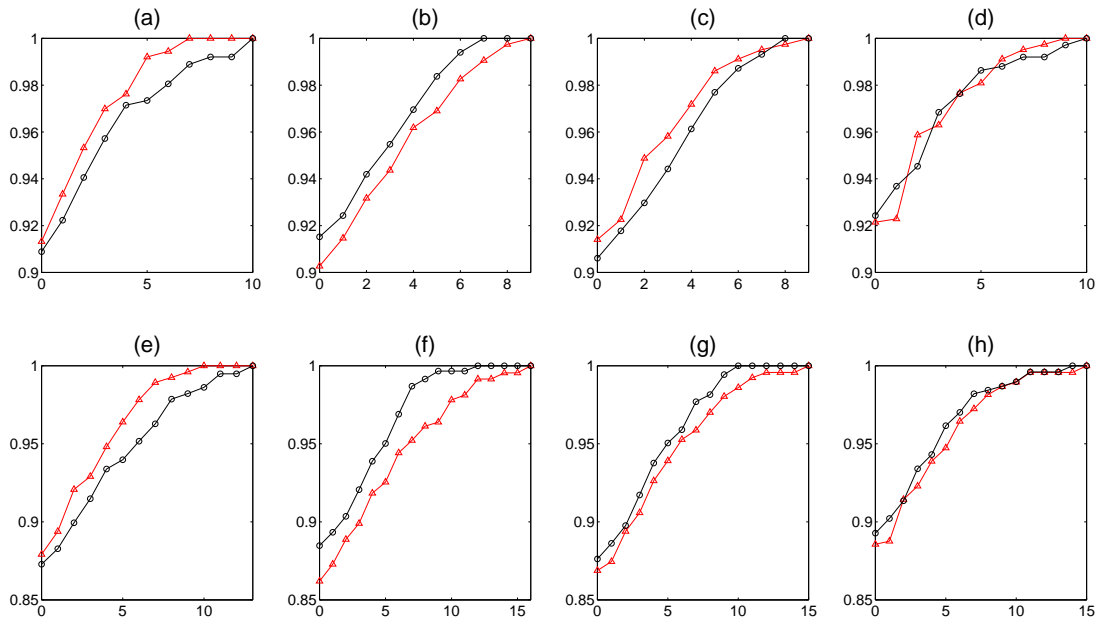}\\
 \caption{\label{fig:distribution}A comparison of the evolutive Kendall's coefficient $\tau_e$ of the evolutive competitivity graph $G^e_c({\mathcal R})$ in terms of the threshold applied for for the German Bundesliga (panels (a) and (e)), the Italian Lega Serie A (panels (b) and (f)), the Spanish Liga BBVA (panels (c) and (g)) and British Premier League (panels (d) and (h)) during 2011-12 (lines in red in all the panels) and 2012-13 (lines in black in all the panels). The evolutive competitivity graphs have been considered at the end of the first part of the season (panels from (a) to (d)) and at the end of the season (panels from (e) to (h)).}
\end{figure*}

Following this idea, Figure~\ref{fig:distribution} shows the fluctuation of the evolutive Kendall's coefficient $\tau_e$  of the {\sl filtered} evolutive competitivity graph $G^e_c({\mathcal R})|_{\ge \omega}$ as a function of the threshold $0\le w\in\Z$. In each panel we have considered the {\sl filtered} evolutive competitivity graph $G^e_c({\mathcal R})|_{\ge \omega}$ for a fixed league (during 2011-12 and 2012-13) and we have computed its evolutive Kendall's coefficient $\tau_e$ as a function of the threshold $0\le w\in\Z$. As before, we have considered the German Bundesliga (panels (a) and (e)), the Italian Lega Serie A (panels (b) and (f)), the Spanish Liga BBVA (panels (c) and (g)) and British Premier League (panels (d) and (h)) during 2011-12 (lines in red in all the panels) and 2012-13 (lines in black in all the panels) in Figure~\ref{fig:distribution}. In order to get more information, we have computed the original evolutive competitivity graph $G^e_c({\mathcal R})$ at two instances in order to get a more detailed information about the competitiveness: at the end of the first part of the season (panels from (a) to (d)) and at the end of the season (panels from (e) to (h)). Since in section~\ref{sec:basic} we have said that a league $L_1$  is more competitive than $L_2$ (with respect to the evolutive Kendall's coefficient) if the evolutive Kendall's coefficient of the evolutive competitivity graph of $L_1$ is less that the corresponding value for the evolutive competitivity graph of $L_2$, the later the increase of the fluctuation of the evolutive Kendall's coefficient occurs, more  competitive a league is. Hence, Figure~\ref{fig:distribution} shows that the four leagues exhibit different behaviors. The competitiveness of the Bundesliga was more competitive in 2012-13 than in 2011-12, either after the first half of the season (panel (a)) and at the end of the season (panel (e)). The competitiveness of the Lega Seria A was less competitive in 2012-13 than in 2011-12, either after the first half of the season (panel (b)) and at the end of the season (panel (f)), i.e. this league has the opposite behavior than the Bundesliga.  The competitiveness of the Liga BBVA was higher in 2012-13 than in 2011-12 after the first half of the season (panel (c)), but it was weaker in 2012-13 than in 2011-12 at the end of the season (panel (g)), so the Spanish league lost competitiveness from one season to the following. Finally, the Premier League kept the strength of the competitiveness in a similar level during seasons 2011-12 and 2012-13, either after the first half of the season (panel (d)) and at the end of the season (panel (h)). Similar analysis can be performed by considering other structural parameters (such as the normalized mean strength, the relative size of the maximal clique and other), as it is illustrated in \cite{suplementary}. In next section we apply some classical measures of competitive balance to the leagues just considered.

\section{Comparison with some results obtained by using standard measures of competitive balance}\label{sec:classic}

The most common measure of competitive balance is the standard deviation of the proportion of wins $w_i$ each team has achieved in a season. This standard deviation can be defined as
\begin{equation}\label{sigma}
\sigma  = \sqrt{\frac 1n \sum_{i=1}^{n} \left(w_i - \frac 12\right)^2},
\end{equation}
where $n$ is the number of teams.
In general, the higher is $\sigma$ the lower is the {\sl competitive balance} (and the higher is the {\sl competitive inequality}). In a league where each team has the same probability of win a match one will have $\sigma =0$, and therefore an increase in $\sigma$ means a decrease in the {\sl competitive balance}. It is said that a league (or, in general, $r$ rankings) has increased its competitiveness when it has decreased its $\sigma$.  Some other measures are variations of this one, see  \cite{Hum},  \cite{KrTh}. In particular, to use as a comparison with our measures, we shall compute the coefficient called NAMSI (National Measure of Seasonal Imbalance, \cite{Goo}):

\begin{equation}\label{namsi}
\hbox{NAMSI}= \sqrt{\frac{\sum_{i=1}^{n} \left(w_i - \frac 12\right)^2}{\sum_{i=1}^{n} \left(w_{i,max} - \frac 12\right)^2}},
\end{equation}
which is a normalized $\sigma$. The term $w_{i,max}$  is the win ratio of team $i$ when there is complete
predictability: team $1$ wins all the matches, team $2$ wins all except two matches (home and away) , ..., team $n$ loses all the matches. In {\sf MATLAB} notation, the vector with components $w_{i,max}$ is computed as
$[n-1:-1:0]/(n-1)$.

Another usual measure of competitive balance is the Herfindahl-Hirschman Index (see \cite{OwRyWe}). A related index, called HICB ($H$-index of competitive balance), see   \cite{Paw}, is defined as
\begin{equation}\label{hicb}
\hbox{HICB} = 100\,n \sum_{i=1}^n s_i^2,
\end{equation}
where  $s_i$ is the ratio of points scored in a season by team $i$. An increase in the HICB coefficient means a decrease in the competitive balance \cite{Paw}, \cite{DoGo}.

A different measure of competitive balance, which is more related with the mathematical literature on rankings,
is the {\em Churn} \cite{Churn}
\begin{equation}\label{churn}
C_t =\frac{1}{n} \sum_{i=1}^{n} \left| c_{i,t} - c_{i,t-1} \right|,
\end{equation}
where $c_{i,t}$ is the ranking of team $i$ in season $t$. Then the {\em Adjusted Churn} is defined as
\begin{equation}\label{adchurn}
C_{ad} =  \frac{C_t}{C_{t,max}},
\end{equation}
where $C_{t,max}$ is the maximum value of $C_t$ given a league of $n$ teams. This happens e.g., when $c_{t}=[1,2,\cdots,n]$ and $c_{t+1}=[n,n-1,...,2,1]$.  This index indicates competitive balance between seasons. If two rankings are the same then $C_t=0$. It is considered that high values of $C_t$ imply high values of
competitive balance \cite{Churn}, \cite{DoGo}.   Note that the Adjusted Churn is a normalized Spearman's footrule $D(c_t,c_{t+1})$. In fact, $D(c_t,c_{t+1})=n C_t$. Standard tools such as the Spearman's footrule
and Kendall's $\tau$ (see equation (\ref{kendalltau})) have also been used in sports ranking (see \cite{Groo}, \cite{DoGo}, \cite{KrTh}). Kendall's $\tau$ allows for a dynamic study of the leagues, in allowing the comparison of two seasons. Following \cite{Groo} a competition is perfectly balanced in a dynamic sense when the ranking is random, that is, any ranking is equally probable, and rankings in consecutive seasons are independent of each other. It is assumed that low values of $\tau$ correspond to high competitive balance.

In table \ref{ta:namsihicb} we show the NAMSI coefficient given by equation (\ref{namsi}), and the HICB coefficient given by (\ref{hicb}). We have computed these coefficients using the data from \cite{web-bundes}, \cite{web-calcio},  \cite{web-liga}, and \cite{web-premier}. According both the NAMSI coefficient and the HICB in the season 2011-12 the most competitive league was Lega Serie A. This is in accordance with our conclusions using $\tau_e$ in section \ref{sec:soccer}. Regarding the season 2012-13 the most competitive league was the Spanish Liga BBVA (by using NAMSI coefficient) or Lega Serie A (by using HICB). Note that the values os these coefficients are very similar for these two leagues in that season. We have obtained, using $\tau_e$ that the most competitive league in this season is Bundesliga. Therefore, in this case our measures give a different conclusion.

\begin{table}
\begin{tabular}{|c|c|c|c|c|}
\hline
   & \multicolumn{2}{|c|}{Season 2011-12}  & \multicolumn{2}{|c|}{Season 2012-13} \\
\cline{2-5}
League     &  NAMSI  & HICB &NAMSI  & HICB \\
\hline
Lega Serie A   & 0.6410 & 107.82 & 0.6495 & 110.40 \\
Liga BBVA      & 0.6581 & 109.74 & 0.6419 & 110.73\\
Premier League & 0.6669 & 110.54 & 0.7135 & 111.33\\
Bundesliga     & 0.6788 & 109.58 & 0.6784 & 111.15\\
\hline
\end{tabular}
\caption{NAMSI and HICB coefficients for the four major European soccer leagues.}
\label{ta:namsihicb}
\end{table}

In table \ref{ta:churn-tau} we show the values of the Adjusted Churn $C_{ad}$, computed by using equation (\ref{adchurn}), and the Kendall's $\tau$ for the four analyzed leagues. Both coefficients $C_{ad}$ show that the Bundesliga is the one that has increased more its competitiveness from season 2011-12 to season 2012-13, since this league has the highest value of $C_{ad}$ and the lowest value of $\tau$. This is in accordance with our conclusions obtained in section \ref{sec:soccer} from panel e) of Figure 3, and from panel e) in Figure 4.  A similar analysis could be done using another classical measures of competitive balance. As we have noted before, depending on whether measure is used one can obtain different conclusions. In section \ref{sec:soccer} we have exploited some of the possibilities of the structural measures introduced in this paper. Some additional results can be obtained from \cite{suplementary}.

\begin{table}
\begin{center}
\begin{tabular}{|c|c|c|}
\hline
  & \multicolumn{2}{|c|}{Season 2011-12 to season 2012-13}  \\
\cline{2-3}
League & Adjusted churn & Kendall's $\tau$ \\
\hline
Lega Serie A   & 0.28  &  0.63\\
Liga BBVA      & 0.34  &  0.54 \\
Premier League & 0.30  &  0.59\\
Bundesliga     & 0.37  &  0.49\\
\hline
\end{tabular}
\caption{Adjusted Churn coefficient, $C_{ad}$, and Kendall's $\tau$ for the four major European soccer leagues.}
\label{ta:churn-tau}
\end{center}
\end{table}

 \section{Conclusions}\label{sec:conclusion}

We have shown that the use of a graph ({\em competitivity graph}) associated to a family of rankings can be useful to analyze the competitiveness associated to the rankings. We have introduced some parameters (mean degree, mean strength, clustering coefficient) that can be used to measure the competitive balance of a sport ranking. We have related the {\em competitivity graph} with classical measures of rankings, such as the
Kendall's correlation coefficient, and we have given a generalized Kendall's correlation coefficient for $r$ rankings that can be computed easily from the competitivity graph.
To show the applicability of our measures we have analyzed the four  major European soccer leagues, obtaining among others the following conclusions:

\begin{itemize}
\item In 2011-12 season the most competitive league was the Italian Lega Serie A,  if we take into account the evolutive Kendall's coefficient, the normalized mean strength and the clustering coefficient (see Table~\ref{table:data12-13}).
\item In 2012-13 season the most competitive league was the German Bundesliga,  if we take into account the evolutive Kendall's coefficient, the normalized mean degree, and the normalized mean strength, see Table~\ref{table:data12-13}).
\item From season 2011-12 to season 2012-13 the league that has increased more its competiteness was the  Bundesliga.
\end{itemize}

\section*{Acknowledgements}

This paper was partially supported by Spanish MICINN Funds and FEDER Funds MTM2009-13848, MTM2010-16153 and MTM2010-18674, and Junta de Andalucia Funds FQM-264.



\end{document}